\documentclass[twocolumn,aps,prl,showpacs,superscriptaddress]{revtex4}
\usepackage{amssymb}

\usepackage{graphicx}

\begin{document}
\title{ Security of decoy-state quantum key distribution
with inexactly controlled source}
\author{Xiang-Bin Wang}
\affiliation{Department of Physics, Tsinghua University, Beijing
100084, China} \affiliation{Imai-Project, ERATO-SORST, JST, Daini
Hongo White Building, 201, 5-28-3, Hongo, Bunkyo, Tokyo 113-0033,
Japan Department of Physics, Tsinghua University, Beijing 100084,
China}
\author{Cheng-Zhi Peng}
\affiliation{Department of
Physics, Tsinghua University, Beijing 100084,
China}\affiliation{Hefei National Laboratory for Physical Sciences
at Microscale and Department of Modern Physics, University of
Science and Technology of China, Hefei, Anhui 230026, China}
\author{Jun Zhang}
\affiliation{Hefei National Laboratory for Physical Sciences at Microscale and
Department of Modern Physics, University of Science and Technology
of China, Hefei, Anhui 230026, China}
\author{Jian-Wei Pan}
\affiliation{Department of Physics, Tsinghua University, Beijing 100084,
China}
\affiliation{Hefei National Laboratory for Physical Sciences
at Microscale and Department of Modern Physics, University of
Science and Technology of China, Hefei, Anhui 230026,
China}
\affiliation{Physikalisches Institut, Universit\"at
Heidelberg, Philosophenweg 12, 69120 Heidelberg, Germany}

\begin{abstract}
The existing theory of decoy-state quantum cryptography assumes the
counting rates of the same state from different sources to be the
same, given whatever channel. This is correct if the intensity of
each individual pulse is controlled exactly. We show by an explicit
example that the assumption is in general incorrect if the source is
inexactly controlled and the error pattern is known to Eve.  We then
further develop the theory of decoy-state method so that it is
unconditionally secure given whatever error pattern provided that
the error is not too large and the bounds of errors are known. 
Our result is not only limitted to the coherent states. It allpies to all
source states satisfying Eq.(17).
\end{abstract}


\maketitle


{\em Introduction.---\/}  Most of the existing set-ups of quantum
key distribution (QKD)\cite{BB84,bruss,GRTZ02,DLH06,GLLP04} use
imperfect single-photon source. Such an implementation in principle
suffers from the photon-number-splitting attack \cite{PNS,PNS1}. The
decoy-state method \cite{rep, H03, Wang05, LMC05,HQph} and some
other methods \cite{scran,kko,zei} can be used for unconditionally
secure QKD even Alice only uses an imperfect source\cite{PNS,PNS1}.

 The separate theoretical
results of ILM-GLLP \cite{GLLP04} shows that a secure final key can
be distilled even though an imperfect source is used in the
protocol, if one knows  the lower bound of the fraction of those raw
bits generated by single-photon pulses from Alice. The decoy-state
method is to verify such a bound faithfully and efficiently.
Recently, a number of experiments on decoy-state QKD have been done
\cite{Lo06,peng}.  However, the existing decoy-state theory assumes
the perfect control of light intensity. This is an impossible task
for any real set-up in practice.
Here we study the  decoy-state method with intensity errors and we
conclude that after revising, the decoy-state method is still secure
and efficient. Our result immediately applies to all existing
experimental results.
\\{\em  Existing theory .---\/}
 Given a class of $N_x$ pulses,
after Alice sends them out to Bob, if Bob observes $n_x$ counts at
his side, the {\em counting rate} for pulses in this class is
\begin{equation}s_x=n_x/N_x.\label{def}\end{equation}
{\bf Proposition 1}: If class $X$ is divided into $l$ subclasses and
any pulse in $X$ belongs to only one subclass and  the fraction of
pulses in each subclasses are $a_0,a_1\cdots,a_l$, then the counting
rate of class $X$ is $ S_X=\sum_{k=0}^l a_ks_k $ and $s_k$ is the
counting rate of the $k$th subclass.
\\
\\{\bf Proposition 2}.
The counting rate of pulses from class $y$ must be equal to that of
class $y'$ if: 1) each class contains sufficiently large number of
pulses; 2)
 pulses of each class are independent and identical;
3) the density operator of a pulse from class $y$ is equal to that
of class $y'$, 4) pulses of $y$ and $y'$ are randomly mixed.


Alice has 3 different sources $Y_0,Y_\mu,Y_{\mu'}$ which can produce
states of $\rho_0=|0\rangle\langle 0|, \rho_\mu,\rho_{\mu'}$ only.
  Suppose
\begin{equation}\label{rho}
\rho_\mu=\sum_{k=0}a_k|k\rangle\langle
k|;\,\rho_{\mu'}=\sum_{k=0}a_k'|k\rangle\langle k|
\end{equation}
and we request the following constraints for all $n>2$:
\begin{equation}\label{ac}
\frac{a_k'}{a_k}>\frac{a_2'}{a_2}.
\end{equation}
Here the states are subscribed by $\mu$ and $\mu'$ because the
diagonal states are actually determined by the intensity $\mu$ or
$\mu'$. For example, a coherent state of intensity $q$ with phase
randomization is
\begin{equation} \rho_\mu =
\sum_{k=0}^\infty \frac{\mu^k e^{-\mu}}{k!}|k\rangle\langle
k|.\label{coh}
\end{equation}
If Alice uses coherent states of intensity 0, $\mu$, and $\mu'$ for
source $Y_0,Y_\mu$ and  $Y_{\mu'}$ in the protocol, the state
parameters are $a_k=\frac{\mu^k e^{-\mu}}{k!}$, $a_k'=\frac{{\mu'}^k
e^{-\mu'}}{k!}$.

We regard the pulses from each source as a {\em class}, i.e.class
$Y_0$ for intensity 0, class $Y_\mu$ for intensity $\mu'$ and class
$Y_{\mu'}$ for intensity $\mu'$. We shall also call the pulses in
class $Y_\mu,~Y_{\mu'}$ the {\em decoy pulses} and the {\em signal
pulses}, respectively for simplicity.  We regard those decoy or
signal pulses containing $k$ photons as (sub-)class $y_k$ or $y_k'$.
We also use the term sub-source for sub-class.

 Proposition 1 leads to the
following relation,
\begin{equation}\label{countdecoy}
S_\mu =\sum_{k=0}^{\infty}a_ks_k=a_0s_0+a_1s_1+\lambda
\end{equation}
and $\lambda=\sum_{k=2}^{\infty}a_ks_k$,  $S_\mu$ is the total
counting rate of all decoy pulses, $s_k$ is the counting rate of
pulses in class $y_k$.  Also, we have
\begin{equation}\label{countsignal}
S_{\mu'}=\sum_{k=0}^{\infty} a_k's_k'
\end{equation}
and  $S_{\mu'},~s_k'$  are the counting rates of all signal pulses
and   class $y_k'$, respectively\cite{Wang05}. In the protocol,
Alice chooses each source randomly in producing each pulse for Bob.
If each source  always produces the assumed state exactly, pulses
from sub-sources $y_0,y_0',Y_0$ are randomly mixed, and so are
classes $y_k,y_k'$. According to Proposition 2,
\begin{equation}\label{0011cc}
s_0=s_0'=S_0,~s_k=s_k'
\end{equation}
where $S_0$ is the counting rate of class $Y_0$. Also, using the
assumed condition in Eq.(\ref{ac}), we can simplify
Eq.(\ref{countsignal}) as
\begin{equation}\label{deff}
S_{\mu'}=\sum_{k=0}^{\infty}a_k's_k\ge
a_0's_0+a_1's_1+\frac{a_2'}{a_2}\lambda .
\end{equation}
According to the definition of counting rate,
 $S_0,S_\mu,S_{\mu'}$ can be observed directly in the protocol.  Since there are only two unknown
variables ($s_1,~\lambda$) for two constraints in
Eq.(\ref{countdecoy}) and Eq.(\ref{deff}), the lower bound value of
$s_1$ can be verified by solving these joint constraints.
\begin{equation}\label{f11}
s_1\ge \frac{a_2'( S_\mu-a_0s_0)-a_2 (
S_{\mu'}-a_0's_0)}{a_2'a_1-a_1'a_2}.
\end{equation}
 Given this, one can calculate the final key rate
by\cite{GLLP04,LMC05}
\begin{equation}\label{ilm}
R=\Delta_1'[1-H(t_1)]-H(t)
\end{equation}
where $\Delta_1'=\frac{a_1s_1}{S_{\mu'}}$ is the fraction of
single-photon counts among the raw bits due to signal pulses and
$t_1$, $t$ are the QBER for single-photon pulses and the QBER for
all signal pulses. This is the result the decoy-state method with
whatever diagonal source-state, including the coherent state,
thermal state, heralded single-photon state, and so on with the
condition of Eq.(\ref{ac}) and exactly controlled source.
\\{\em The consequence of correlated intensity error.---\/}The
existing  theory assumes that the parameters for state of {\em each
individual decoy pulses and signal pulses} keep to be constant.
 In practice, these parameters cannot be constant for all pulses\cite{wang07,wangapl}.
   If the parameter errors  are independent,
   then we can \cite{wang07} use the averaged state for  each
source to verify the lower bound of $s_1$ by Eq.(\ref{f11}).
However, the issue becomes more complicated in the most general
situation when there are correlated intensity errors. As  shown
explicitly below, in such a case, the average-state method fails and
we have to consider the instantaneous intensity of each individual
pulses.

 There can be correlated intensity
errors which are not known to Alice and Bob but is known to Eve.
(This situation is more realistic if one uses the decoy-state
plug-and-play protocol\cite{gisind}, where Eve. can actually prepare
the error pattern). Consider a specific example. In the whole
protocol, the pulses are divided into $M$ blocks. Each block
contains $N/M$ pulses and $N=N_0+N_\mu+N_{\mu'}$, where
$N_0,N_\mu,N_{\mu'}$ are number of pulses from source
$Y_0,Y_\mu,Y_{\mu'}$, $N$ is the total number of pulses. The state
of each individual pulse from source $Y_0,Y_{\mu'}$ are always
controlled exactly. But there are errors for the state of pulses
from source $Y_\mu$ (decoy pulse). Say, in half of those $M$ blocks,
state for every decoy pulse  is actually vacuum (we name any of such
blocks as block $D_0$).  In the other half of blocks, the intensity
of each decoy pulse is twice of the assumed value, i.e., $2\mu$, and
$\mu$ is the value for the assumed intensity that Alice {\em wants}
to use for each decoy pulse. We name these blocks as block $D_2$.
Note that either block $D_0$ or block $D_2$ consists of three
classes, $Y_0,Y_\mu,Y_{\mu'}$. Eve. knows such type of error pattern
but Alice does not know it. We shall assume $\mu=0.2$ and $\mu'=0.6$
in the protocol. Actually, by watching the averaged photon number of
each block, Eve can know exactly the intensity of decoy pulses in
each block, i.e., 0 or $2\mu=0.4$, provided that $N/M$ is not so
small.

Here is Eve's scheme using {\bf time-dependent channel}: she blocks
all pulses from block $D_0$, and she produces a linear channel of
transmittance $\eta_e$ to attenuate each pulse from block $D_2$.
Straightly, the actual counting rate for the sub-source $y_1$ is
$s_1=\frac{2\eta_e\mu e^{-2\mu}}{2\mu e^{-2\mu}}= \eta_e$ and the
actual counting rate for sub-source $y_1'$ is
\begin{equation}\label{s1p}s_1'=\frac{1}{2}\frac{\eta_e\mu'e^{-\mu'}}{\mu'
e^{-\mu'}}=\eta_e/2.\end{equation} Obviously, $s_1\not=s_1'$.
Similarly, we can also show that $s_k\not=s_k'$. This shows, given
the correlated error which is known to Eve, Eve can treat the pulses
from sub-sources $y_k$ and $y_k'$ {\em differently} ! Proposition 2
of the existing decoy-state theory can not be used anymore.  Given
the correlated intensity error, pulses from sub-source $y_k$ and
$y_k'$ are actually {\em not randomly mixed}: in some blocks, the
number of $k-$photon decoy pulses are larger than that of other
blocks.

 If one disregards the fact that Proposition 2 does not hold now
 and go ahead to use Eq.(\ref{f11}) with the {\em averaged state} for
the decoy pulse, the protocol will be insecure because the
calculated value of the single-photon counting rate by the existing
theory will be larger than the true value. For simplicity, we assume
$S_0=0$. Given the Eve's scheme above,  Alice and Bob will find
\begin{equation}
S_{\mu}=\frac{1-e^{-2\eta_e\mu}}{2}\approx 0.2\eta_e;
~S_{\mu'}=\frac{1-e^{-\eta_e\mu'}}{2}\approx 0.3\eta_e
\end{equation}
 If we go ahead to assume $s_k=s_k'$ and blindly use
Eq.(\ref{f11}),  the counting rate of sub-source $y_1'$ can be
obtained by replacing $a_i$ with the $0.4^ke^{-0.4}/k!/2$ there. The
result for the counting rate of single-photon pulses by
Eq.(\ref{f11}) is
\begin{equation}
s_1=s_1'\ge 2.65336\eta_e
\end{equation}
which is larger than the real value $\eta_e/2$ as shown
Eq.(\ref{s1p}). This means, {\em the value of counting rate for
single-photon from signal pulses verified by the protocol would be
larger than the true value} and hence the protocol is insecure.

A more realistic case\cite{rep} that in certain blocks, intensities of all pulses
are a bit higher and in other blocks intensities of all pulses are a bit lower than the assumed
values, it is also found that $s_1\not= s_1'$ therefore existing decoy state theory cannot
be used blindly.  
\\ {\em Remark on proposition 2.---\/}
   The above study has clearly shown that, if there are correlated intensity
   errors of light intensity,
 Proposition 2 cannot be blindly  used because here $y_1$ and $y_1'$ are {\em not}
 randomly mixed and the result is insecure given the time-dependent Eve's channel.

 {\em Our solution.---\/}
In the actual protocol, each pulse sent from Alice is randomly
chosen from one of 3 sources $\{Y_0,Y_\mu,Y_{\mu'}\}$ with
probability $p_0,p_\mu,p_{\mu'}$ ($p_0+p_\mu+p_{\mu'}=1$). For
simplicity, we assume that every pulse in class $Y_0$ is exactly in
vacuum state. But each single shot of pulse in classes
$Y_\mu,Y_{\mu'}$ can be in a state slightly different from the
expected one. We assume at time $i$ Alice actually produces
\begin{equation}\label{rhoi}
\rho_{\mu i}=\sum a_{ki}|k\rangle\langle k|
\end{equation}
or
\begin{equation}\label{rhoip}
\rho_{\mu'i}=\sum a_{ki}'|k\rangle\langle k|
\end{equation}
instead of $\rho_\mu$ or $\rho_{\mu'}$ as defined in Eq.(\ref{rho}),
where the parameters of each $|k\rangle\langle k|$ are constant. If
Alice uses coherent states, the time-dependent parameters
$a_{ki},a_{ki}'$ are determined by the time-dependent intensities by
$a_{ki}=\frac{\mu_i^k e^{-\mu}}{k!},~a_{ki}'=\frac{\mu_i'^k
e^{-\mu_i'}}{k!}$, and $\mu_i$( or $\mu_i'$) is the instantaneous
intensity of the $i$'th pulse ( decoy  signal pulse). Lets first
consider a virtual protocol, {\bf Protocol 1}: At each time $i$ in
sending a pulse to Bob, Alice produces a two-pulse (pulse $A$ and
pulse $B$) bipartite state
\begin{equation}\label{r2}
\rho_i^2=p_0 |z_0\rangle\langle z_0|\otimes|0\rangle\langle 0|+p_\mu
|z_1\rangle\langle z_1|\otimes \rho_{\mu i} + p_{\mu'} |z_2\rangle\langle
z_2|\otimes \rho_{\mu_i'}
\end{equation}\
 Here the first subspace is for pulse $A$ and
the second subspace is for pulse $B$;
 $i$ runs from 1 to $N$, the total number of pulses
transmitted to Bob. States $\{|z_x\rangle\}$ are orthogonal to each
other for different $x$ ($x=0,1,2$). Alice keeps pulse $A$ and sends
out pulse $B$ to Bob.
      After Bob completes the detection, Alice measures  pulse $A$.  The outcome of
$|z_0\rangle,|z_1\rangle$ or $|z_2\rangle$ of pulse $A$ corresponds
to class $Y_0,Y_\mu$ or $Y_{\mu'}$ for pulse $B$. Asymptotically,
the number of pulses in theses classes should be
$N_0=p_0N,~N_\mu=p_\mu N,~N_{\mu'}=p_{\mu'}N$.  We use notation
$a_k^L,a_k^U$ for lower bound and upper bound of
$\{a_{ki}|i=0,1\cdots N\}$, $a_k'^L,a_k'^U$ for lower bound and
upper bound of $\{a_{ki}'|i=0,1,\cdots N\}$. Also, we need
\begin{equation}\label{con2}
\frac{a_k'^L}{a_k^U}\ge \frac{a_2'^L}{a_2^U}>1
\end{equation}
for all $k> 2$.
 In the protocol, Alice {\em could} have known the
exact photon number in any pulse before sending it to Bob. We define
those pulses containing $k$ photons as class $\tilde y_k$, e.g., all
the single-photon pulses make of class $\tilde y_1$.  For clarity,
we also define set $\{l_k\}$ ($k=0,1,2\cdots$): for the $i$th pulse
($i$ from 1 to $N$), if the pulse contains $k$ photons, then $i\in
l_k$. Since Alice could have known the photon number in each pulse,
any $i$ must belong to only one $l_k$. Asymptotically, there should
be $N_k=\sum_1^{N}(p_\mu a_{ki}+p_{\mu'}a_{ki}')$ pulses (elements)
in class (set) $\tilde y_k$ ($l_k$) for any $k>0$ and there are
$N_0=\sum_{i=1}^N(p_0+a_{0i}p_\mu+a_{0i}'p_{\mu'})$ pulses (elements
) in class (set) $\tilde y_0$ ($l_0$). Also, each individual pulse
can only belong to one class from $\{\tilde y_k|k=0,1,\cdots \}$.
Most generally, we assume the {\em instantaneous} transmittance of
the $i$'th pulse  to be $\eta_{i}$ ($i$ from 1 to $N$), which should
be either 0 or 1. If the pulse causes no count at Bob's side, the
instantaneous counting rate for that pulse is 0, otherwise it is 1.
Given any photon number state $|k\rangle\langle k|$, it can be from
both class $Y_\mu$ and class $Y_{\mu'}$. Asymptotically, the numbers
of counts caused  by decoy pulses (pulses from class $Y_\mu$),
signal pulses and pulses in class $Y_0$ are
\begin{equation}
n_\mu=\sum_{i\in l_0} p_\mu a_{0i}d_{i}+\sum_{k=1}^\infty\sum_{i\in
l_k}{p_\mu a_{ki}d_{i}},
\end{equation}
\begin{equation}
n_{\mu'}=\sum_{i\in l_0} p_{\mu'}
a_{0i}'d_{i}+\sum_{k=1}^\infty\sum_{i\in l_k}{p_{\mu'} a_{ki}'d_{i}}
\end{equation}
and
\begin{equation}\label{ndk}
n(Y_{0})=\sum_{i\in l_0}\frac{p_0}{p_0+p_\mu a_{0i}+p_{\mu'}a_{0i}'}
\eta_{i},
\end{equation}
respectively, where $d_i$ is defined as
\begin{eqnarray}
d_{i}=\frac{\eta_{i}}{p_\mu a_{0i}+p_{\mu'}a_{0i}'+p_0};\; {\rm
if}\;\;i\in l_0
\\
d_{i}=\frac{\eta_{i}}{p_\mu a_{ki}+p_{\mu'}a_{ki}'};\; {\rm
if}\;\;i\in l_k,\;k\ge 1
\end{eqnarray}
 The values of $n_\mu,n_{\mu'}$ and $n(Y_0)$ can be directly
observed in the protocol, therefore are regarded as known
parameters. Based on Eq.(\ref{ndk}), we have the following
inequality: \begin{equation}\sum_{i\in l_0} p_\mu a_{0i}d_i \le
\frac{p_\mu a_0^U n(Y_0)}{p_0}=p_\mu a_0^U S_0N,
\end{equation}
\begin{equation}p_{\mu'}a_{0i}'d_i \ge \frac{p_{\mu'}
a_0'^L n(Y_0)}{p_0}=p_{\mu'} a_0'^L S_0N
\end{equation}
and $S_0=\frac{n(Y_0)}{p_0N}$, the counting rate of class $Y_0$,
which is observed in the protocol itself. Given bounded values for
parameters $\{a_{ki},a_{ki}'\}$, we have
\begin{equation}
n_{\mu}\le p_\mu a_0^U S_0N+ p_\mu a_{1}^U\sum_{i\in l_1} d_{i}
+p_\mu\sum_{k=2}^\infty a_k^U(\sum_{i\in l_k} d_{i})
\end{equation}
and
\begin{equation}
n_{\mu'}\ge p_{\mu'} a_0'^L S_0N + p_{\mu'} a_{1}'^L\sum_{i\in l_1}
d_{i} +p_{\mu'}\sum_{k=2}^{\infty} a_k'^L(\sum_{i\in l_k} d_{i}).
\end{equation}
Using the assumed conditions given in Eq.(\ref{con2}) we can solve
the above simultaneous constraints and obtain
\begin{equation}
 \frac{1}{N}\sum_{i\in l_1} d_{i} \ge \frac {a_2'^L
S_{\mu}-a_2^U
S_{\mu'}+a_0'^La_2^US_0-a_0^Ua_2'^LS_0}{a_2'^La_1^U-a_1'^La_2^U}
\end{equation}
where $S_\mu = \frac{n_\mu}{p_\mu N}=\frac{n_\mu}{N_\mu}$,
$S_{\mu'}=\frac{n_{\mu'}}{N_{\mu'}}$.   The fraction of
single-photon counts among those counts caused by signal pulses is
$\frac{\sum_{i\in l_1} a_{1i}' d_{i}}{p_{\mu'}NS_{\mu'}}$, which is
lower bounded by
\begin{equation}\label{delta1'}
\Delta_1' \ge \frac{a_1'^L(a_2'^L S_{\mu}-a_2^U
S_{\mu'}+a_0'^La_2^US_0-a_0^Ua_2'^LS_0)}{S_{\mu'}(a_2'^La_1^U-a_1'^La_2^U)}
\end{equation} and the fraction of single-photon counts of  decoy pulses
is lower bounded by
\begin{equation}
\Delta_1  \ge \frac{a_1^L(a_2'^L S_{\mu}-a_2^U
S_{\mu'}+a_0'^La_2^US_0-a_0^Ua_2'^LS_0)}{S_{\mu}(a_2'^La_1^U-a_1'^La_2^U)}
\end{equation}
 For coherent states, if the intensity is bounded by
$[\mu^L,\mu^U]$ for decoy pulses and $[\mu'^L,\mu'^U]$ for signal
pulses then
\begin{equation}
a_k^{X}=(\mu^X)^ke^{-\mu^X}/k!,~a_k'^{X}=(\mu'^X)^ke^{-\mu'^X}/k!
\end{equation}
with $X=L,U$ and $k=1,2$.  Therefore,  one can calculate the final
key rate  by Eq.(\ref{ilm}) now, if the bound values of intensity
errors  are known. The asymptotic result using the experimental data
of QKD distance of 50 kilometers calculated by our formula is listed
in table I.


\begin{table}
\caption{\label{tab:table1}Secure key rate ($R$) vs different values
of intensity error upper bound ($\delta_M$) using the experimental
data in the case of 50~km~\cite{peng}. The experiment lasts for
1481.2 seconds with the repetition rate 4~MHz. We have observed
$S_{\mu'}=3.817\times 10^{-4}, S_{\mu}=1.548\times 10^{-4},
S_{0}=2.609\times 10^{-5}$ and the  quantum bit error rates (QBER)
for signal states and decoy states are $4.247\%, 8.379\%$
respectively. }
\begin{ruledtabular}
\begin{tabular}{lllllll}
 $\delta_M$ & $5\%$ & $4\%$ & $3\%$ & $2\%$ & $1\%$ & 0\\
\hline
$R$ (Hz) & 70.8 & 84.3 & 97.6 & 110.7 & 123.6 & 136.3\\
\end{tabular}
\end{ruledtabular}
\end{table}

The results above are for the virtual protocol where Alice uses the
bipartite state of Eq.(\ref{r2}). Obviously Alice can choose to
measure all A-pulses  of each bipartite state in the very beginning
and the virtual protocol is reduced to the normal protocol in
practice, where the bipartite state is not necessary.
\\{\em Application in the Plug-and-Play protocol.---\/}
 As shown by Gisin et al\cite{gisind}, combining with the decoy-state method,
 the plug-and-play protocol can be unconditionally secure.
  There, Alice receives strong pulses from Bob
and she needs to guarantee the exact intensity of the pulse sending
to Bob. It is not difficult to monitor the intensity, but difficult
to {\em control} the intensity. Our theory here can help to save the
difficult feed-back intensity control: Alice monitors the
intensities, discards those pulses whose intensity error is too
large, and then use our theory with the known bound bound of
intensity errors.
\\{\em Concluding remark:} In summary, we have shown the the unconditional security
of decoy-state method given what-ever error patern of the source, provided that the parameters
diagonal
state of the source satisfy Eq.(17) and the bound values of each parameters in the state is known.
Our result here applies to whatever distribution of source state that satisfies Eq.(17). Our result also
answers clearly the often asked question ``What happens if the state of Laser beam is not exactly in the 
assumed distribution ?". Here we don't need exact information about the source state, we only need the
bound values for parameters $a_1,a_2.1_1',a_2'$ and Eq.(17).
 \\{\bf Acknowledgement:} This work was
supported in part by the National Basic Research Program of China
grant No. 2007CB907900 and 2007CB807901, NSFC grant No. 60725416 and
China Hi-Tech program grant No. 2006AA01Z420.

\end{document}